# Development of a hardened THz energy meter for use on the kilojoule-scale, short-pulse OMEGA EP laser

G. Bruhaug,[1,b)] H. G. Rinderknecht,[1] Y. E,[2] M. S. Wei,[1] R. B. Brannon,[1] D. Guy,[1] R. G. Peck,[1] N. Landis,[1] G. Brent,[1] R. Fairbanks,[1] C. McAtee,[1] T. Walker,[1] T. Buczek,[1] M. Krieger,[1] M. H. Romanofsky,[1] C. Mileham,[1] K. G. Francis,[2] X. C. Zhang,[2] G. W. Collins,[1] and J. R. Rygg[1]

[1]*Laboratory for Laser Energetics, University of Rochester, Rochester, NY, 14623-1299 USA*
[2]*The Institute of Optics, University of Rochester, Rochester, NY, 14627 USA*

A highly adaptable and robust THz energy meter has been designed and implemented to detect energetic THz pulses from high-intensity ($>10^{18}$ W/cm$^2$) laser–plasma interactions on OMEGA EP. THz radiation from the laser driven target is detected by a shielded pyrometer. A second identical pyrometer is used for background subtraction. The detector can be configured to detect THz pulses in the 1-mm to 30-$\mu$m (0.3- to 10-THz) range and pulse energies from joules to microjoules via changes in filtration, aperture size and position. Additional polarization selective filtration can also be used to determine THz pulse polarization. The design incorporates significant radiation and EMP shielding to survive and operate within the OMEGA EP radiation environment. We describe the design, operational principle, calibration and testing of the THz energy meter. The pyrometers were calibrated using a benchtop laser and show linear sensitivity up to 1000 nJ of absorbed energy. Initial results from four OMEGA EP THz experiments detected up to ∼15 $\mu J$ at the detector, which can correspond to 100's of mJ depending on THz emission and reflection models.

## I. INTRODUCTION

Terahertz (THz) radiation interacts with matter in neither a purely photonic nor a bulk electronic fashion.[1–3] Due the unique nature of THz radiation, there is a large interest in high-power sources for nonlinear time-domain spectroscopy and relativistic light–matter interactions at the extremes of low frequency.[1,3] However, generation of such THz pulses is extremely difficult with traditional methods. Recent work with laser-plasma THz generation has shown great promise in scaling THz pulses to the TW and >100-mJ scale using short-pulse (≤ps), J to kJ-scale lasers to drive solid, liquid, or gaseous targets.[1] To maximize the THz power and pulse energy, lasers with both high energy (100's of J to kJ class) and high intensity ($>10^{18}$ W/cm$^2$) must be used. These lasers are most commonly single shot and are well known for their immense electromagnetic pulse (EMP),[4] hard x-ray,[5] and charged-particle generation.[6] The OMEGA EP laser is especially challenging due to the peak EMP field measured being one of the highest seen on any laser ($>10^2$ kV/m)[4]. This adds to the already challenging task of THz detection due to the low efficiency (average of ∼0.1%) of laser THz generation in these systems. All available THz detection methods rely on electronics[2] rather than passive methods, such as film, further compounding the EMP noise issue in these experiments.

This paper outlines the development of a ten-inch manipulator (TIM)-mounted THz energy meter for use on the kilojoule-class OMEGA EP laser and the associated challenges with the development of this detector. The THz background/energy meter (TBEM) is a broadband (0.3- to 10-THz or 1-mm to 30-$\mu$m) energy meter based on THz-sensitive pyrometers and capable of detecting broadband THz pulses as weak as ~50 $\mu$J emitted in $4\pi$ or as strong as ~2 J emitted in $4\pi$ before suffering saturation of the detection element.

a)The author to whom correspondence should be addressed: gbru@lle.rochester.edu

## II. DIAGNOSTIC OVERVIEW

### A. Mechanical design

TBEM is a 112.5-cm-long, 20.9-cm-wide TIM-mounted diagnostic weighing 33.1 kg largely due to the 19.8 kg of tungsten radiation shielding. The use of a TIM for detector mounting allows for TBEM to be inserted into the OMEGA-EP chamber at a variable depth. As shown in Fig. 1, the diagnostic consists of a light-tight aluminum chassis with a front-mounted TPX[7] (THz and optical light transmissive) lens and filter pack extending 36.8 cm from the main body. This front lens allows for THz radiation to be collected 15 cm from target chamber center (TCC) when the TIM is fully inserted, maximizing the sensitivity of the detector. In front of the lens is a removable high-resistivity silicon[7] (HRFZ-Si) wafer that acts as a THz-transmissive blast shield. A front filter pack can hold THz filters to alter the portion of the spectrum sampled as well as irises to reduce the amount of THz radiation transmitted down the optical path. This lens and filter assembly can also be removed and the detector operated retracted from the target chamber, and with the aperture at 116 cm from TCC, to further protect the electronics from EMP and radiation.

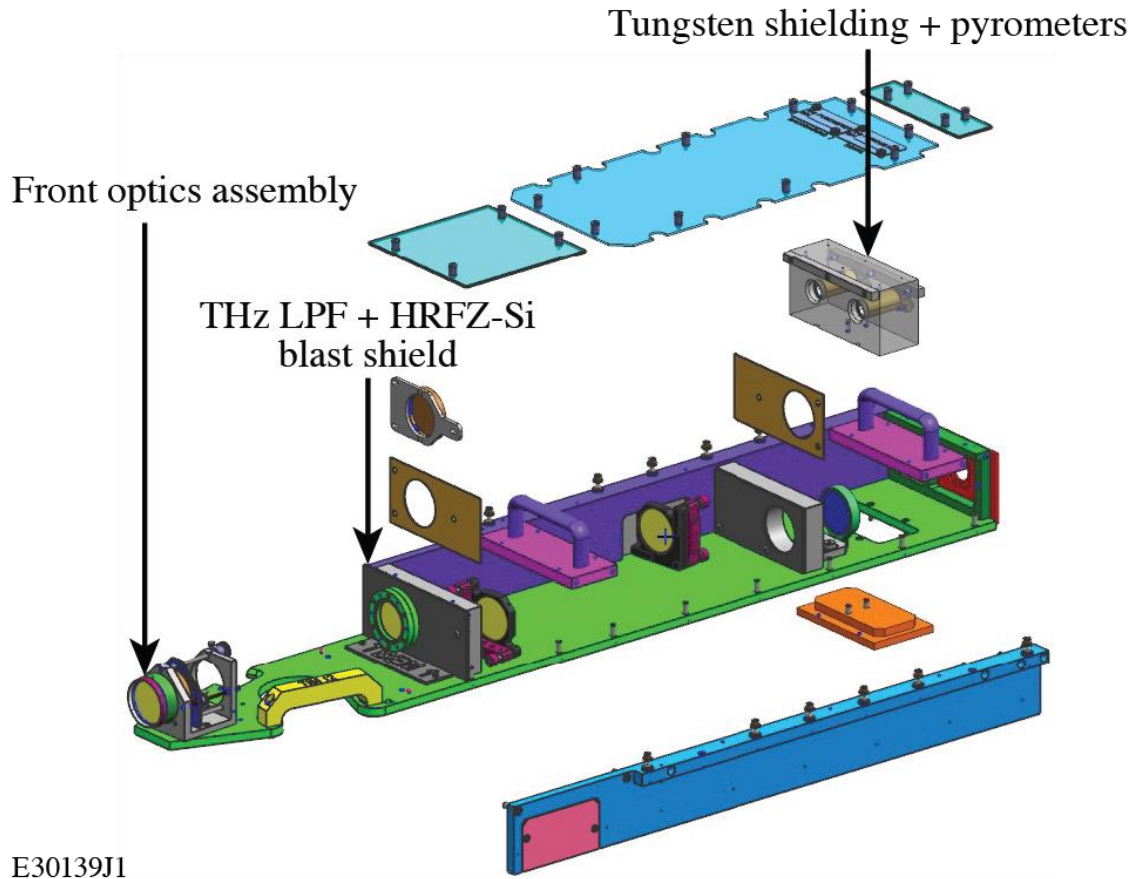

**FIG. 1.** Expanded view of TBEM detector assembly.

The front lens assembly collimates the THz radiation source and sends it though the 5.4-cm aperture in the first tungsten shielding block shown in Fig. 1. A HRFZ-Si wafer and 10.9-THz low-pass filter[7] (LPF) prevent IR, optical, and UV light from passing through this aperture. The collimated THz light then reflects off of two gold-coated mirrors and passes through a second tungsten 5.4-cm aperture to a TPX lens that refocuses the radiation onto the THz-sensitive pyrometer through a 1-mm thick piece of polytetrafluoroethylene (PTF)[7] acting as a second low-pass THz filter placed directly onto the pyrometer. A second THz pyrometer sits 7.3 cm off to the side of the first pyrometer and is covered with a 1-mm thick aluminum plate to block THz so as to act as a background radiation detector. Both pyrometers are placed within a 7.6-cm-thick tungsten block for radiation shielding. The full optical path and THz transmission of various components can be seen in Fig. 2. The spectrally averaged THz transmission factor through the full TBEM optics without any extra filters is 0.144, but can be increased to 0.278 without the use of the front lens and filter pack.

**B.  THz detection**

THz detection is accomplished with the use of commercially available nanojoule sensitive pyrometers,[8] which are commonly used for commercial and scientific THz detection. A pyrometer is a broadband-sensitive energy meter that relies on the pyroelectric effect to detect a change in energy deposition.[2,8] The sensor is built in a series of layers similar to a capacitor with two electrodes around an inner layer of pyroelectric material. One electrode is darkened to best absorb the wavelength range of interest. Upon pulsed irradiation, the pyrometer will heat up and the polarization direction in the pyroelectric material will change. A charge disparity then develops across the pyroelectric crystal and a voltage pulse is generated that is proportional in amplitude to the amount of energy deposited onto the sensor. The decay of the signal can vary based on local thermal conditions and is not relevant for the detection of THz pulses that are shorter than the response time of the detector, which is in microseconds[9]. An outline of the sensor and the resulting physics of operation are shown in Fig. 3.

Since these sensors are broadband and extremely sensitive, care must be taken to ensure that only THz radiation reaches the detector.

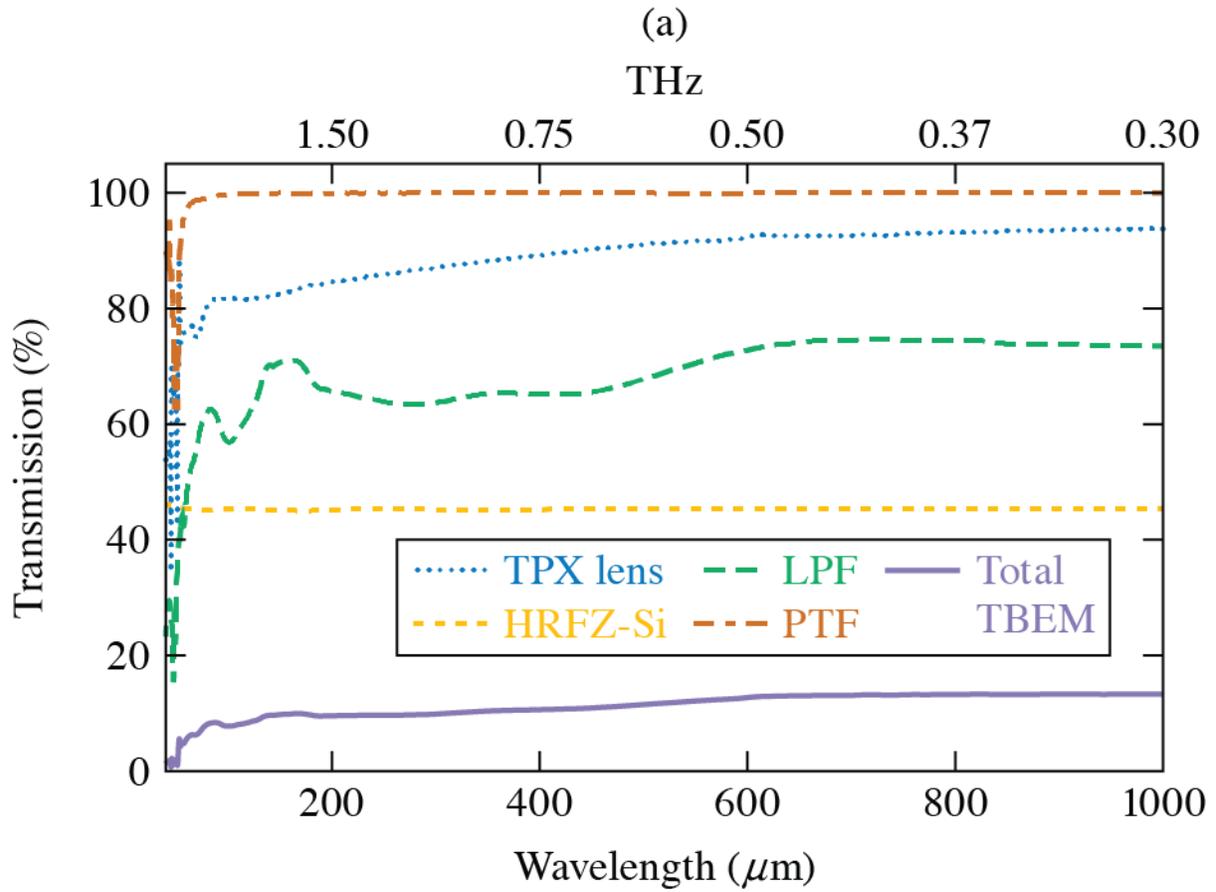

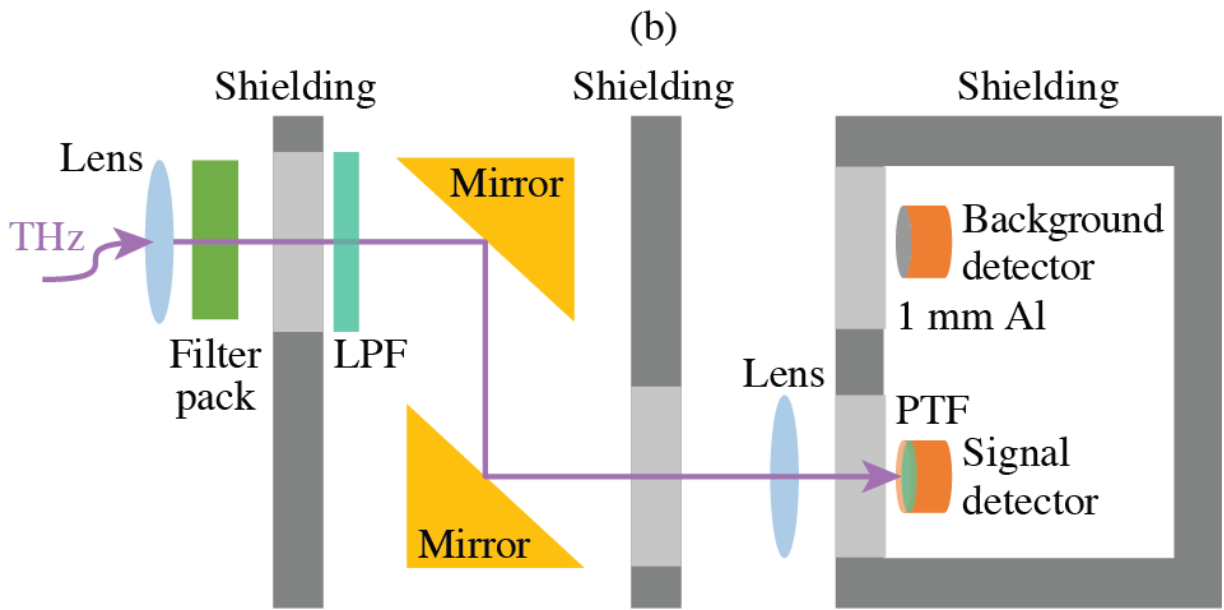

**FIG. 2.** **(a)** THz transmission of optical components and full detector[7,8](data courtesy of Gentec Electro-Optics, Inc. www.gentec-eo.com and Tydex LLC, www.tydexoptics.com). **(b)** Full optical path of the TBEM detector.

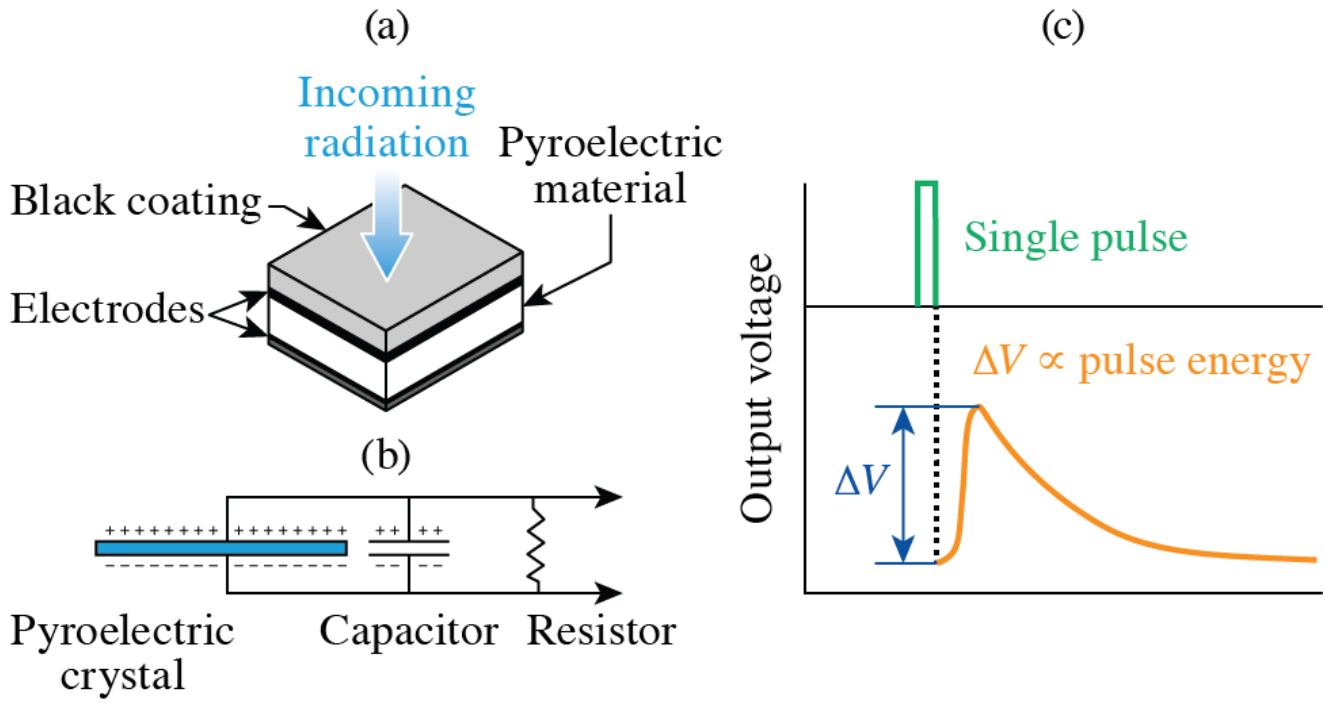

**FIG. 3. (a)** A diagram of a typical pyrometer and **(b)** a basic circuit diagram of a pyrometer. (c) Typical voltage response versus input radiation pulse.

### C. EMP and radiation shielding

During short-pulse laser operation, EMP fields in excess of $10^2$ kV/m have been measured on OMEGA EP.[4] To protect the electronics of TBEM, either BNC or twisted pair cables are used for all external signal paths, and these are wrapped in one layers of copper foil, two layers of copper mesh, and two layers of aluminum foil. The case was sealed with aluminum tape to prevent any light and or RF leaks. A microwave mesh was tested at the front 5.4 cm aperture, but found to not affect the noise seen during experiments while having an unacceptably high THz attenuation and subsequently was removed from further experiments. The detectors are also always operated under battery power, so as to reduce any chance of induced signal from the AC power cable or building power.

Short-pulse laser experiments also generate prodigious x-ray and charged-particle fluxes. Both 10's of MeV-class electrons[5] and MeV-class x rays[6] have been measured on OMEGA EP and can induce erroneous signals in pyrometers if not properly shielded. The shielding shown in Fig. 1 and Fig. 2(a) can reduce MeV electron and x-ray flux by greater than a factor of 10,000 (Ref. 9).

### D. Detector calibration and benchtop testing

Both TBEM detectors built were tested with a cw THz laser to confirm their responses to THz radiation. They were tested for light tightness with a 600-W tungsten bulb and had no response from any angle due to the filtration blocking all ultraviolet, optical, infrared, and mid-infrared light. Benchtop EMP testing was performed using the 12-J, 700-fs Multi-Terawatt (MTW) laser as well as a 60-kV Tesla coil. Three layers of foil cable shielding and an EMP suppression circuit were added (Fig. 4).

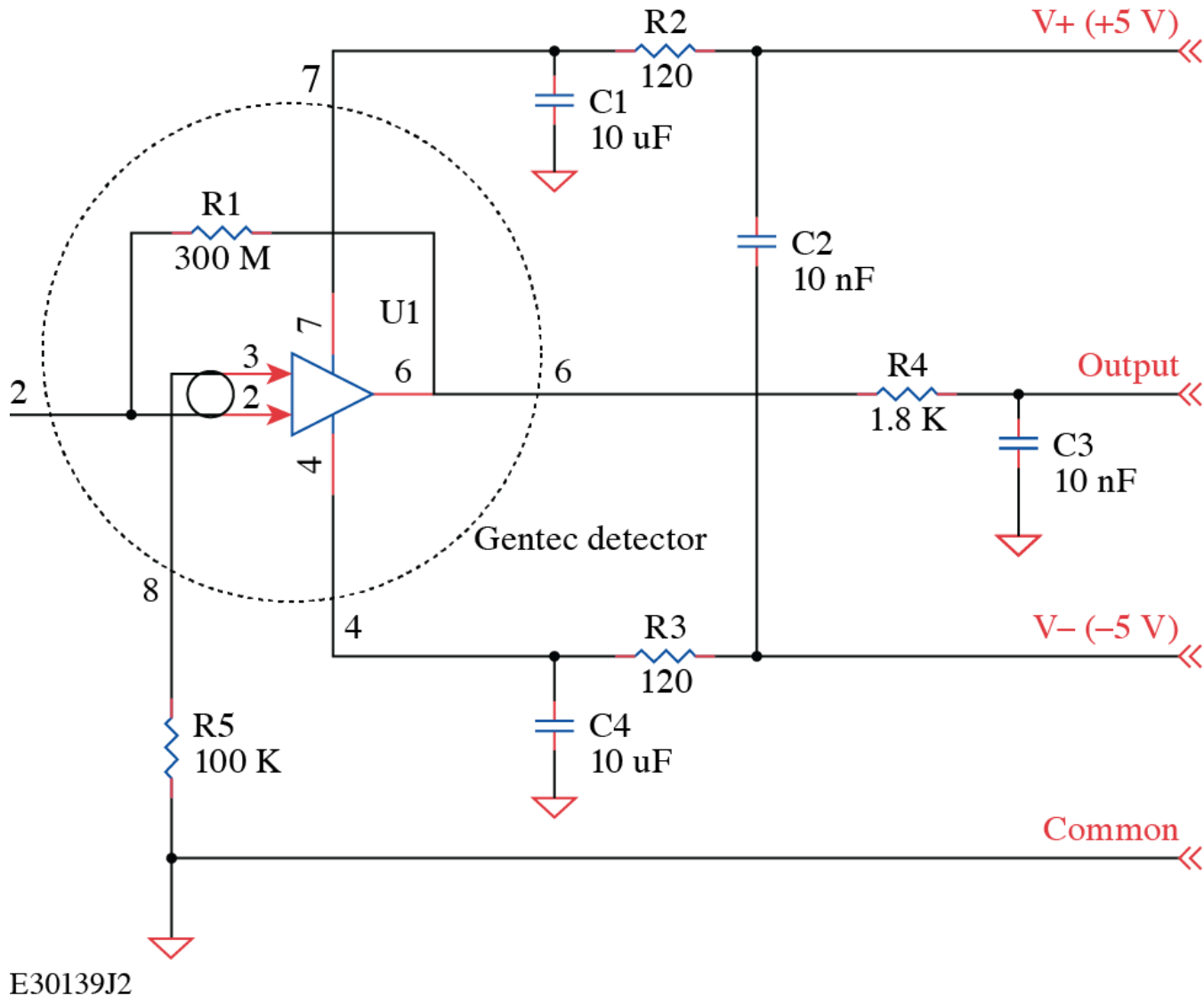

**FIG 4.** Pyrometer circuit with EMP suppression circuit.

All pyrometers were energy calibrated using a mJ-class 1-$\mu$m laser operated with 10-ps pulses in single-shot mode. The laser energy was varied from 10's of nJ to ~4 $\mu$J using a variable optical attenuator. A beam splitter was used to send ~1/30th of the laser energy to an already calibrated laser energy meter. The peak voltage of the signal was then cross-calibrated to the energy on the detector and calibration plot [Fig. 5(a)] were made for all four pyrometers. The detectors were observed to be roughly linear up to a laser energy of ~1000 nJ, or ~5000 mV peak signal, above which the sensitivity follows a shallower curve. A model for the sensitivity curve from 0 to 5000 nJ is:

$$E(\text{nJ}) \approx \begin{cases} 0.1439\, S(\text{mV}) & [S < 6000 \text{ mV}] \\ 1.45\, S(\text{mV}) - 7835.7 & [S \geq 6000 \text{ mV}] \end{cases}$$

The results from 1-$\mu$m light can be adjusted for THz radiation using a manufacturer provided calibration curve [Fig 5(b)][8] and in general the pyrometers are ~10× more sensitive to 1 $\mu$m than to THz radiation. An example 1-$\mu$m energy calibration signal can be seen in Fig. 6(a). It was found that all pyrometers responded roughly the same.

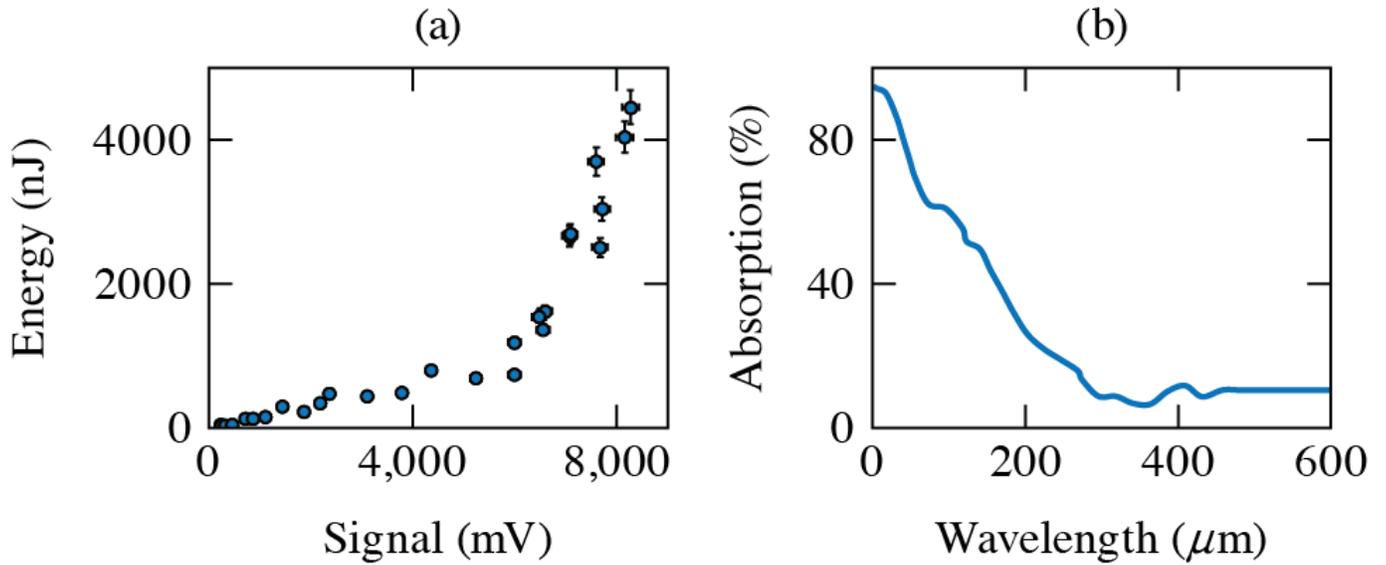

E30139J3

**FIG. 5.** (a) Energy calibration results for all four pyrometers using a 1-$\mu$m calibrated laser and (b) the absorption curve of the pyrometers versus wavelength from optical to THz.[8] (data courtesy of Gentec Electro-Optics, Inc. www.gentec-eo.com). The manufacturer states that the detectors continue to have a flat absorption out to 1 mm or greater, but did not provide the calibration data for this.

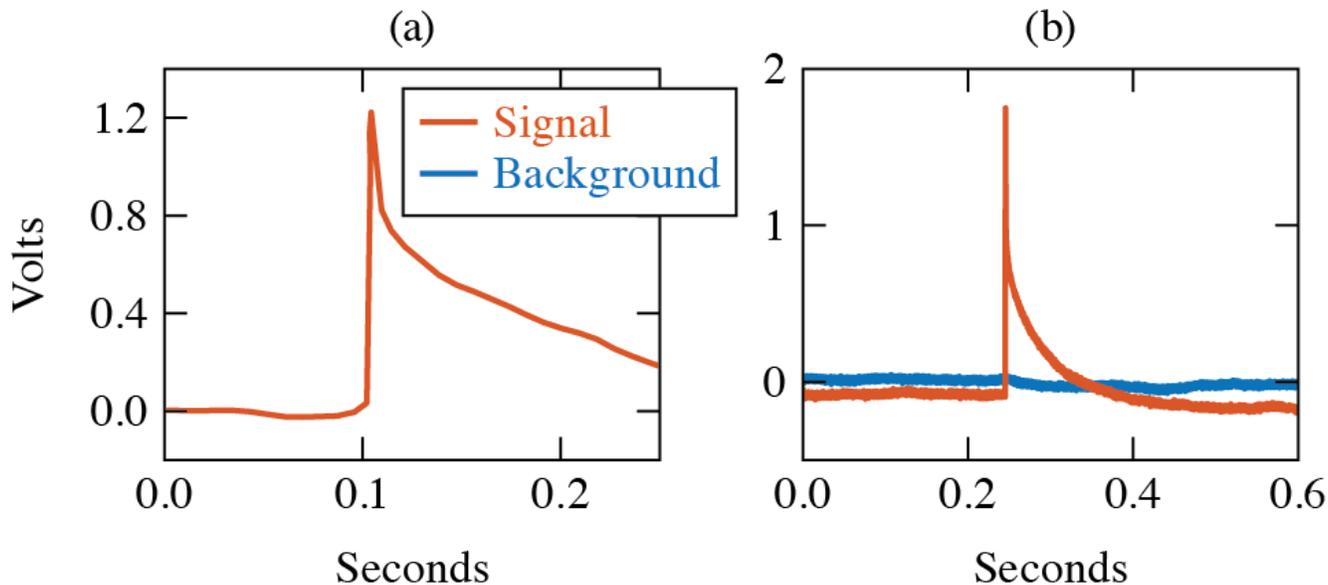

**FIG. 6.** Example signal trace from ~170 nJ, 1-$\mu$m calibration laser pulse (a) and TBEM analogue on MTW (b) with two pyrometer channels for THz signal and background radiation respectively. The difference is signal position is an artifact of oscilloscope triggering.

### III. INITIAL DETECTOR TESTING

#### A. Detector concept tests on MTW

The initial concept for TBEM was built and tested for use on MTW to support THz target design development. The results have been extremely promising, as shown by an example THz waveform from MTW compared to a 1-$\mu$m laser calibration in Fig. 6.

The large signal seen in Fig. 6(b) is indicative of the very strong (mJ, GW-class) THz signals expected from J-class, high-intensity laser–plasma THz generation experiments[1]. One important finding from the MTW experiments is the use of plastic target stalks to reduce EMP. Although this is well-known technique in short-pulse laser experiments,[4] it was found that the noise levels were unacceptable on the pyrometers unless plastic stalks were used.

#### B. Initial performance on OMEGA EP

Four campaigns have been undertaken on OMEGA EP to test the TBEM detectors using best laser compression (700 fsec) with the final two campaigns showing repeatable THz detection. Both detectors were placed to face the rear of the target, which is predicted to generate the maximum THz yield[1]. The first campaign in July 2021 did not have the full complement of EMP and radiation shielding in place and suffered from massive x-ray and EMP noise problems [see Fig. 7(a)]. The detectors were both operated fully inserted for this campaign. As with on MTW, it was found that using plastic target stalks reduced the noise, but in this case the reduction was not enough to reliably observe THz signal above the background level. All targets tested were foils and the laser energy was varied from 50 to 500 J without any success.

Testing using the MTW laser as an x-ray, and later EMP, source indicated that upgrades to radiation and EMP shielding were needed. A second OMEGA EP campaign was then undertaken using foil targets with more promising initial results and once again used the detectors fully inserted. Figure 7(b) shows an example signal from the second campaign where a small potential THz signal can be seen, but the signal was still unacceptably noisy.

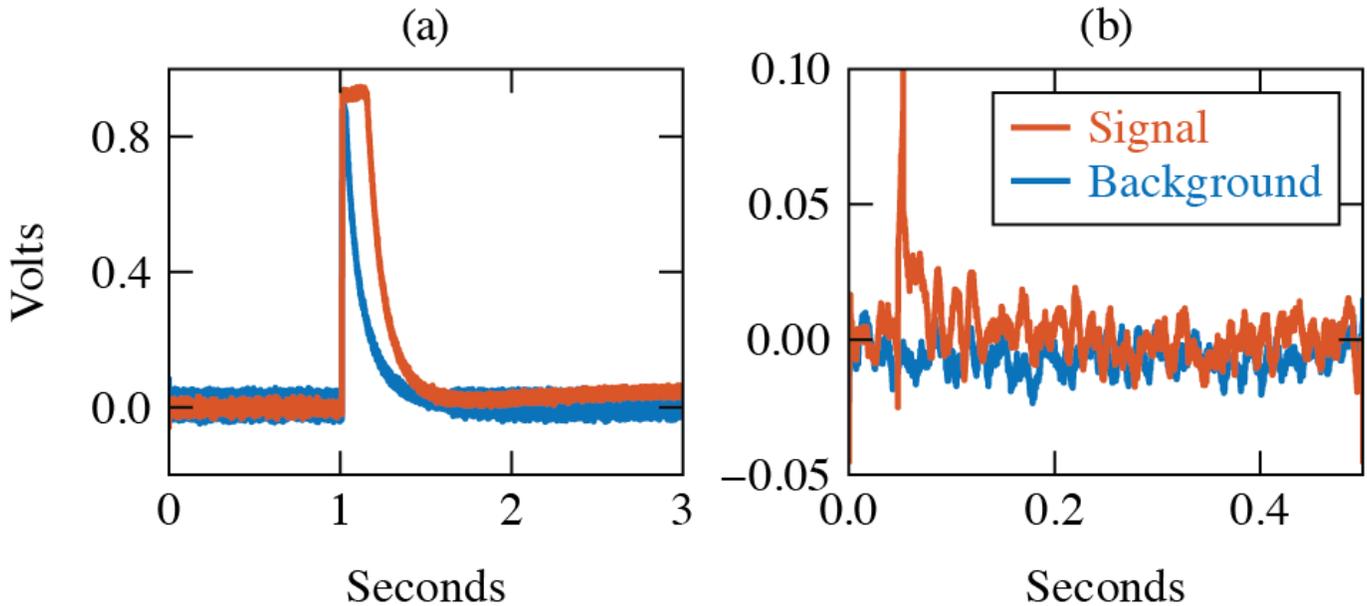

E30140J1

**FIG. 7. (a)** An example of data taken from one TBEM on the first OMEGA EP THz laser-foil campaign in July 2021. **(b)** Data taken on the second OMEGA EP THz campaign with ~100 J of laser energy in March 2022 with a 5-mm iris installed to further reduce THz fluence as a device safety precaution. Extremely highly background noise attributed to EMP and x-rays can be seen in **(a)** when there was less shielding present on the detector, while **(b)** is noisier than would be preferred.

The potential THz signal reduced to nothing after the addition of the aforementioned microwave absorbing mesh to the beam path of both TBEM detectors. After the shot day, the microwave absorbing mesh was tested and found to be highly absorptive in the THz regime. A post-shot day inspection also found damage to the internal EMP shielding and a decision was made to add a further layer of EMP shielding to the wiring between the pyrometers and the detector casing after repairs were undertaken, cumulating in the final three layers of solid metal shielding in the design.

The third and fourth campaigns happened back-to-back on OMEGA EP in June 2022 using two different THz generation target types, foils and microchannels. Example THz detections from these experiments can be seen in Fig. 8. The detectors were operated in both inserted and retracted configurations during these campaigns to test the full range of capabilities of the detectors, while laser energy was varied from 100 J–300 J.

The detectors are estimated to have received ~7 $\mu J$ of THz and ~15 $\mu J$ from the foil and microchannel targets respectively. Depending on assumptions of THz emission angles and chamber reflectivity, this can correspond to total THz energies up to ~150 mJ from the foil target and ~300 mJ from the microchannel target respectively. This is roughly in line with the estimated generation efficiencies of these target types for the given laser energy[1,10]. Total THz yields were estimated using the pyrometer sensitivity curve, the THz absorption factor of the detector (~1/10), the broad band THz transmission (0.278 for the optics used in the detection of the shown signals) and the total THz collection efficiency based on detector position (a factor of ~0.00005 for the experiments shown). It should be noted that these are extremely preliminary estimates that will require full THz emission and reflection models to confirm. A 60 Hz oscillation noise can be seen on top of the THz signal. This is believed to be due to vacuum pump induced vibrations based on additional benchtop testing. The additional EMP shielding on the pyrometer wiring was found to be crucial for the most energetic laser shots. Shots were undertaken to confirm the importance of the additional EMP shielding and a comparison with and without the shielding can be seen in Fig. 9.

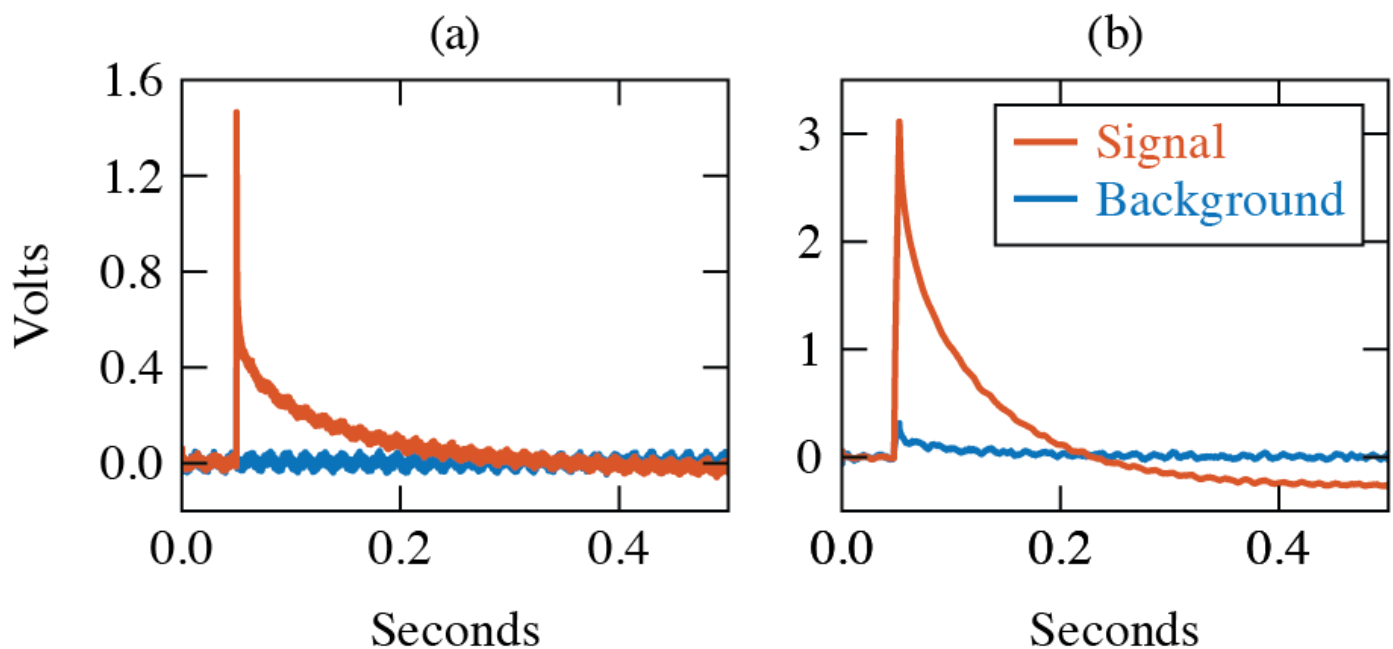

E30512J1

**FIG. 8. (a)** An example of data taken from one TBEM on the third and fourth OMEGA EP campaign (June 2022) coming from a foil irradiated with ~100 J of laser energy. **(b)** A microchannel target irradiated with ~300 J of laser energy.

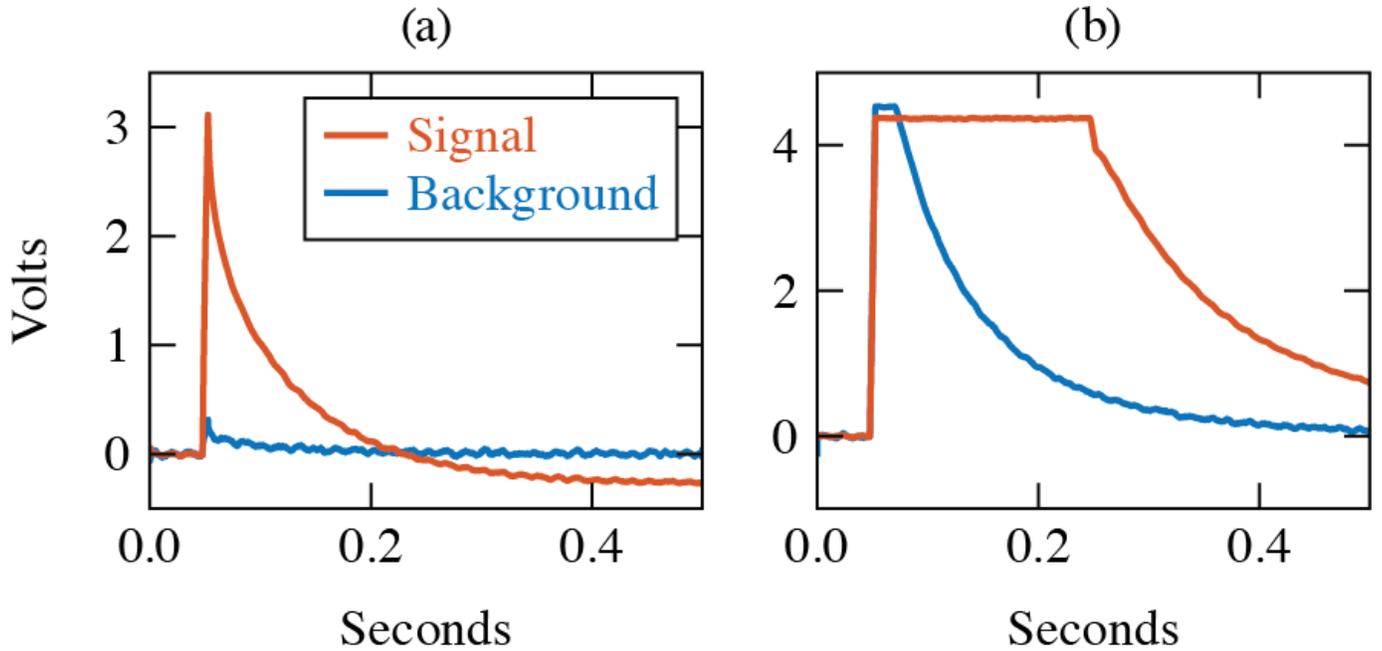

**FIG. 9. (a)** An example of TBEM response with the added EMP shielding and **(b)** without the added shielding. The THz generation source was a microchannel target irradiated with a ~300 J laser in both cases.

The immediate pyrometer saturation and large increased background signal confirmed the utility of these extra layers of EMP shielding. The final design includes three layers of solid metal foil shielding on each stage of signal cable until reaching the power supply of the pyrometers. An additional two layers of mesh are included as both mechanical protection and further EMP shielding.

## IV. DISCUSSION AND FUTURE PLANS

Two broadband THz energy meters were built for use on OMEGA EP during short-pulse experiments and shown to work on the benchtop. The detector concept was then shown to function on the joule-class MTW laser and successfully detected laser–plasma-generated THz radiation. When initially used on the kilojoule-class OMEGA EP laser, however, the detectors suffered excess background from both x rays and EMP. Additional radiation shielding and EMP shielding eventually reduced the background to near zero, although several iterations were needed to achieve this.

For future THz detectors on kilojoule-class, short-pulse lasers, it is recommended that all detector components be placed external to the experimental chamber. This configuration was found to be much more successful on the joule-class MTW laser and made iteration on detector components much faster. The THz radiation should be transmitted as far as possible from the extremely harsh environment of a short-pulse laser-plasma experiment to help reduce background radiation and EMP induced detector noise. Encoding the THz signal into an easier form of radiation to detect, such as the infrared laser used in THz time-domain spectrometry (TDS),[2] would further increase the signal-to-noise ratio (SNR) of these difficult experiments and help transport the signal even further. Future THz detectors on OMEGA EP will utilize THz-TDS techniques to both increase SNR and allow for single-shot spectrometry[2] to take place.

**ACKNOWLEDGMENTS**

This material is based upon the work supported by the Department of Energy, the National Nuclear Security Administration, under Award No. DE-NA0003856, Air Force Office of Scientific Research (FA9550-21-1-0300, FA9550-21-1-0389); National Science Foundation (ECCS-1916068), the University of Rochester, and the New York State Energy Research and Development Authority.

This report was prepared as an account of work sponsored by an agency of the U.S. Government. Neither the U.S. Government nor any agency thereof, nor any of their employees, makes any warranty, express or implied, or assumes any legal liability or responsibility for the accuracy, completeness, or usefulness of any information, apparatus, product, or process disclosed, or represents that its use would not infringe privately owned rights. Reference herein to any specific commercial product, process, or service by trade name, trademark, manufacturer, or otherwise does not necessarily constitute or imply its endorsement, recommendation, or favoring by the U.S. Government or any agency thereof. The views and opinions of authors expressed herein do not necessarily state or reflect those of the U.S. Government or any agency thereof.

The authors thank Gentec-EO and Tydex LLC for their technical expertise and help with the development of these detectors.